\documentclass[conference, a4paper]{IEEEtran}
\IEEEoverridecommandlockouts

\ifCLASSINFOpdf
  \usepackage[pdftex]{graphicx}
  \DeclareGraphicsExtensions{.pdf,.jpeg,.png}
\else
\fi

\usepackage{cite}
\usepackage{amsmath,amssymb,amsfonts,amsthm}
\usepackage{algorithmic}
\usepackage{graphicx}
\usepackage{textcomp}
\usepackage[left=1.4cm, right=1.4cm, top=1.7cm]{geometry}
\usepackage{caption}
\usepackage{anyfontsize}
\theoremstyle{plain}      

\usepackage{cleveref}
\crefformat{equation}{(#2#1#3)}
\Crefname{algocfline}{Algorithm}{Algorithms}
\Crefname{algocf}{line}{lines}
\Crefname{AlgoLine}{Line}{Lines}
\crefname{AlgoLine}{line}{lines}

\makeatletter
\newcommand*\titleheader[1]{\gdef\@titleheader{#1}}
\AtBeginDocument{%
\let\st@red@title\@title
\def\@title{%
\bgroup\normalfont\normalsize\centering\@titleheader\par\egroup
\vskip0.2em\st@red@title}
}
\makeatother

\makeatletter
\renewcommand{\fnum@figure}{Figure \thefigure}
\makeatother

\title{Noncooperative Coordination for \\
Decentralized Air Traffic Management
\vspace{0.5cm}
}

\titleheader{Second US-Europe Air Transportation Research and Development Symposium (ATRDS2026)}

\author{\IEEEauthorblockN{Jaehan Im}
\IEEEauthorblockA{Department of Aerospace Engineering \\
The University of Texas at Austin \\
Austin, Texas, USA \\
jaehan.im@utexas.edu}
}

\renewcommand\IEEEkeywordsname{Keywords}
\IEEEaftertitletext{\vspace{-1\baselineskip}}

\begin{document}
\pagestyle{plain}
\maketitle
\newcommand{\activea}{\mathcal{I}}
\newcommand{\aSet}{{\mathcal{I}}}
\newcommand{\xSet}{\bm{x}}
\newcommand{\activef}{\mathcal{F}}
\newcommand{\fSet}{{\mathcal{F}}}

\newcommand{\ones}{\bm 1}
\newcommand{\reals}{{\mbox{\bf R}}}
\newcommand{\integers}{{\mbox{\bf Z}}}
\newcommand{\symm}{{\mbox{\bf S}}}  % symmetric matrices

\newcommand{\nullspace}{{\mathcal N}}
\newcommand{\range}{{\mathcal R}}
\newcommand{\Rank}{\mathop{\bf Rank}}
\newcommand{\Tr}{\mathop{\bf Tr}}
\newcommand{\diag}{\mathop{\bf diag}}
\newcommand{\card}{\mathop{\bf card}}
\newcommand{\rank}{\mathop{\bf rank}}
\newcommand{\conv}{\mathop{\bf conv}}
\newcommand{\prox}{\bm{prox}}

\newcommand{\Expect}{\mathop{\bf E{}}}
\newcommand{\Prob}{\mathop{\bf Prob}}
\newcommand{\Co}{{\mathop {\bf Co}}} % convex hull
\newcommand{\dist}{\mathop{\bf dist{}}}
\newcommand{\argmin}{\mathop{\rm argmin}}
\newcommand{\argmax}{\mathop{\rm argmax}}
\newcommand{\epi}{\mathop{\bf epi}} % epigraph
\newcommand{\Vol}{\mathop{\bf vol}}
\newcommand{\dom}{\mathop{\bf dom}} % domain
\newcommand{\intr}{\mathop{\bf int}}
\newcommand{\sign}{\mathop{\bf sign}}
\newcommand{\norm}[1]{\left\lVert#1\right\rVert}
\newcommand{\mnorm}[1]{{\left\vert\kern-0.25ex\left\vert\kern-0.25ex\left\vert #1 
    \right\vert\kern-0.25ex\right\vert\kern-0.25ex\right\vert}}

\newtheorem{definition}{Definition} 
\newtheorem{theorem}{Theorem}
\newtheorem{lemma}{Lemma}
\newtheorem{corollary}{Corollary}
\newtheorem{remark}{Remark}
\newtheorem{proposition}{Proposition}
\newtheorem{assumption}{Assumption}
\newtheorem{example}{Example}

\newcommand{\cf}{{\it cf.}}
\newcommand{\eg}{{\it e.g.}}
\newcommand{\ie}{{\it i.e.}}
\newcommand{\etc}{{\it etc.}}

\newcommand{\putref}{{\color{red}[r]}}

\newcommand{\ba}[2][]{\todo[color=orange!40,size=\footnotesize,#1]{[BA] #2}}

\newcommand{\fix}[1]{\textcolor{red}{#1}}

\newcommand{\bigO}{\mathcal{O}}

\newcommand{\intSet}{\mathbb{Z}}
\newcommand{\realSet}{\mathbb{R}}
\newcommand{\natSet}{\mathbb{N}}
\newcommand{\zeroSet}{\bm{0}}
\newcommand{\state}{\bm{x}}

\newcommand{\param}{\kappa}

\newcommand{\plSet}{\mathbf{N}}
\newcommand{\chSet}{\mathbf{M}}
\newcommand{\opSet}{\mathcal{O}}

\newcommand{\xVec}{\bm{x}}
\newcommand{\coordFactor}{\mathbf{w}}

\newcommand{\circnum}[1]{%
  \raisebox{.5pt}{\textcircled{\raisebox{-.9pt}{#1}}}%
}

\noindent \begin{abstract}
Decentralized air traffic management requires coordination among self-interested stakeholders operating under shared safety and capacity constraints, where conventional centralized or implicitly cooperative models do not adequately capture this setting. 
We develop a unified perspective on noncooperative coordination, in which system-level outcomes emerge by designing incentives and assigning signals that reshape individual optimality rather than imposing cooperation or enforcement. 
We advance this framework along three directions: scalable equilibrium engineering via reduced-rank and uncertainty-aware correlated equilibria, decentralized mechanism design for equilibrium selection without enforcement, and structured noncooperative dynamics with convergence guarantees. 
Beyond these technical contributions, we discuss core design principles that govern incentive-compatible coordination in decentralized systems. 
Together, these results establish a foundation for scalable, robust coordination in safety-critical air traffic systems.
\end{abstract}

\vspace{0.3cm}


\section{Introduction} 
Air traffic management (ATM) has traditionally relied on centralized coordination, although operational authority is often distributed across local sectors, adjacent-sector coordination, and network-level traffic flow management \cite{atm_central,atm_central_2,atm_problem_1,atm_problem_2_aam}. 
This centralized structure has provided essential safety and coordination functions, but increasing traffic density introduces scalability challenges \cite{atm_central_2, atm_problem_1}, which are further amplified in emerging traffic paradigms such as urban air mobility \cite{atm_problem_2_aam}. 
Moreover, centralized systems may lack the flexibility and resilience required to adapt to disruptions and rapidly evolving operational environments \cite{cent_resilience_prob, cent_resilience_prob_2}. 

These limitations have motivated the development of decentralized ATM concepts, in which decision-making authority is distributed among multiple stakeholders \cite{atm_central_2, decent_intro, decent_intro_2}. In such settings, airlines, sectors, and regional authorities operate with local objectives, private cost structures, and limited willingness to share sensitive operational information. 
Consequently, system-level performance is no longer dictated by centralized optimization, but instead emerges from strategic interactions among stakeholders. 

A substantial body of research has investigated decentralized ATM architectures from multiple perspectives. Early efforts in distributed air/ground traffic management proposed redistributing decision authority from centralized control centers to aircraft operators and ground systems \cite{decent_intro, decent_intro_2, prev_decent}. Subsequent work explored decentralized air traffic flow management methods based on distributed optimization and multi-agent coordination \cite{prev_decent_nash, prev_decent_block, prev_decent, tomlin, brugnara2023market}. In parallel, research in urban traffic management and low-altitude advanced air mobility management adopted partially decentralized architectures involving multiple service providers \cite{prev_decent_utm, prev_decent_utm_2, prev_decent_block, atm_central_2, prev_decent_utm_hier}. More recently, UAM-oriented ATM architectures have emphasized hierarchical structures to address resilience requirements in emerging urban airspace systems \cite{atm_central_2, prev_decent_utm_hier}. 

Much of the existing literature on conflict resolution and flow management, however, assumes cooperative behavior or central enforcement with explicitly noncooperative formulations appearing only in limited cases \cite{noncoop_required, tomlin}. At the same time, recent work recognizes that, in the absence of a central authority, competition over shared resources must ultimately be resolved through negotiation- or incentive-based mechanisms, yet clear and widely adopted strategic protocols remain lacking \cite{noncoop_required, prev_decent_utm}. These observations suggest the explicit modeling of stakeholder interaction as a noncooperative interaction remains comparatively underexplored. 
% Therefore, decentralized ATM should be modeled as a noncooperative multi-agent system in which stakeholders pursue individual objectives while operating under shared safety and capacity constraints. 

We propose a research topic that investigates the following central question:

\begin{figure}[thpb]
\centering
\framebox{\parbox{3in}{`` How can coordination emerge in decentralized air traffic systems among self-interested, noncooperative stakeholders without relying on a central authority?
 "}}
\label{fig:centralquestion}
\end{figure}

\noindent
We address this question through a unified perspective on noncooperative coordination via incentive design. 
We first summarize the design principles that enable coordination among self-interested stakeholders under shared safety and capacity constraints. 
We then review three interconnected research directions: scalable equilibrium engineering, decentralized mechanism design with convergence guarantees, and structured noncooperative dynamics with system performance guarantees, including dynamic incentive control for large-scale air traffic systems.

\section{Design Principles for Decentralized ATM Coordination}

Before presenting the research program, we summarize several recurring principles that guide the design of decentralized air traffic management (ATM) coordination mechanisms.

\paragraph{Incentive compatibility}

We cannot assume that stakeholders to follow imposed instructions. 
Coordination must therefore be \emph{individually rational}. 
Correlated equilibrium constructs incentive-compatible signals \cite{my_rrce}, TACo resolves preference conflicts without enforcement \cite{my_taco}, and chance-constrained formulations preserve incentive compatibility under cost uncertainty.

\paragraph{Convergence guarantees}

Safety-critical operations require predictable behavior. 
Potential game structures ensure finite-time convergence of best-response dynamics to pure Nash equilibria \cite{my_brd}. 
Oversight-based negotiation further guarantees termination and bounded performance degradation, preventing oscillatory or unstable dynamics \cite{my_iteartive}.

\paragraph{Authority-autonomy trade-off}

Full decentralization may reduce efficiency, while full centralization limits scalability and stakeholder autonomy. 
Parameterized models, such as the cooperativeness factor $\kappa$ \cite{my_brd} and taxation-based oversight \cite{my_iteartive}, quantify intermediate regimes in which calibrated intervention improves system performance while balancing individual decision authority.

\paragraph{Robustness-efficiency trade-off}

Private and uncertain costs necessitate robustness guarantees. 
Confidence parameters in chance-constrained correlated equilibria explicitly trade off deviation risk against achievable efficiency, making robustness a tunable design variable \cite{my_ccce, my_rrccce}.

\paragraph{Scalability}

ATM coordination scales combinatorially with traffic density \cite{my_brd, my_rrce, my_taco}. 
Reduced-rank correlated equilibria restrict coordination to low-dimensional subspaces spanned by multiple Nash equilibria, enabling tractable computation in large action spaces \cite{my_rrce}.

\paragraph{Privacy and fairness}

Stakeholders are unwilling to disclose sensitive cost information. 
Auction-based and equilibrium-based mechanisms preserve private valuations \cite{my_taco, my_iteartive} while improving fairness and security relative to uncoordinated outcomes \cite{my_rrce, my_taco, my_iteartive, my_security}.

\section{Research Program: \\Noncooperative coordination}

We propose three interconnected research areas that has been investigated.

\subsection{Equilibrium engineering and scalable computation}

Equilibrium concepts model how self-interested agents interact in decentralized systems. 
Uncoordinated strategic behavior typically converges to Nash equilibrium, which often yields inefficient and unfair outcomes.

Correlated equilibrium expands the achievable outcome space by allowing a coordinator to send signals that align individual incentives with system-level objectives. 
Rather than predicting behavior, we treat equilibrium as a design variable: the coordinator constructs a distribution over joint actions such that no stakeholder benefits from unilateral deviation.

\begin{definition}[Correlated equilibrium]
Consider a finite game with action sets $A_i$ and utilities $u_i$. 
A distribution $z \in \Delta(A)$ over joint actions $A = \prod_i A_i$ is a correlated equilibrium if
\begin{equation}
\textstyle{
\sum_{a_{-i}} z(a_i,a_{-i}) \big( u_i(a_i,a_{-i}) - u_i(a_i',a_{-i}) \big) \ge 0,
}
\end{equation}
for every player $i$ and any actions $a_i,a_i' \in A_i$.
\end{definition}

While correlated equilibrium enables efficient coordination without centralized enforcement, its direct computation scales exponentially with the number of agents and actions. 
To address this limitation, we developed the Reduced-Rank Correlated Equilibrium (RRCE) method.

\subsubsection{Reduced-rank correlated equilibria \cite{my_rrce}}

RRCE approximates the correlated equilibrium set by restricting coordination to the convex hull of a finite collection of pre-computed Nash equilibria. 
Instead of enumerating all $m^n$ joint actions in an $n$-player $m$-action game, the method operates in a low-dimensional subspace spanned by selected equilibria, dramatically reducing computational complexity.

\begin{figure}[t!]
    \centering
    \includegraphics[width=0.85\linewidth]{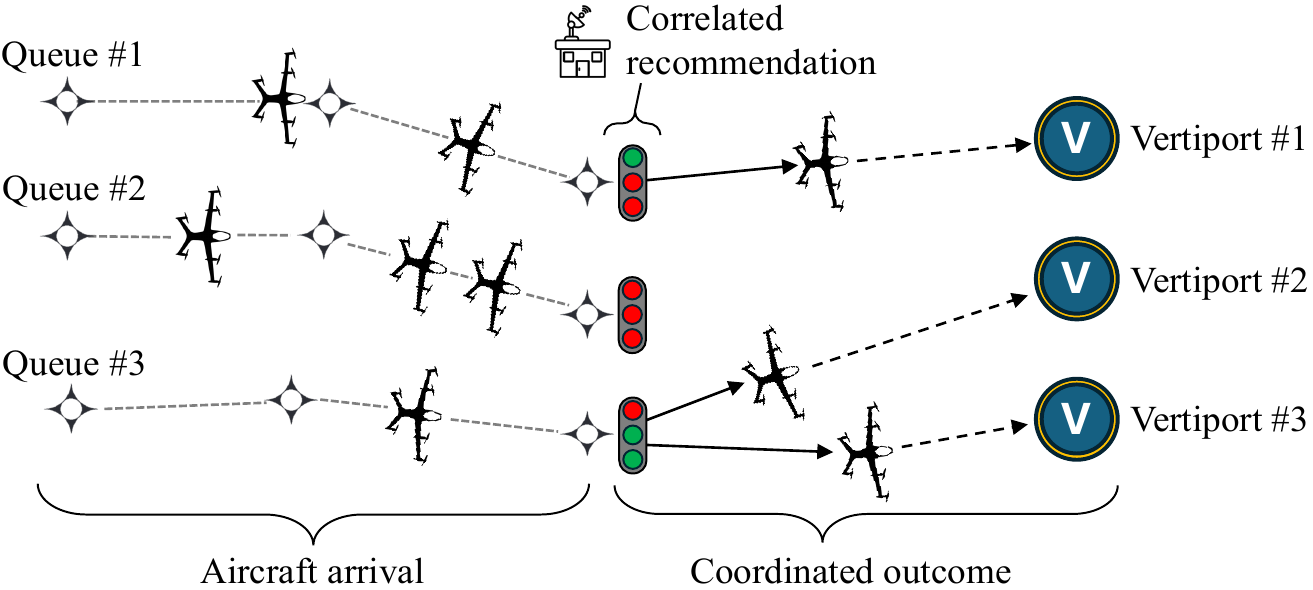}
    \caption{Decentralized vertiport allocation scenario. 
    Multiple airline-specific departure queues compete for limited runway capacity. 
    Each airline selects pushback or runway assignment actions based on its own delay cost, while a coordinator broadcasts correlated signals to improve system-level efficiency. 
    The resulting game models noncooperative competition over shared departure resources.}
    \label{fig:rrce}
\end{figure}

We evaluate RRCE in decentralized vertiport allocation and virtual queue coordination scenarios, illustrated in \Cref{fig:rrce}, where airlines act as self-interested agents competing for shared departure capacity. 
Numerical experiments show that RRCE achieves near-centralized performance while preserving autonomy and incentive compatibility. 
Compared to Nash equilibrium outcomes, RRCE significantly reduces accumulated delay and improves fairness across airlines, while maintaining scalability to large action spaces that are otherwise intractable for direct correlated equilibrium computation.

\subsubsection{Uncertainty-aware correlated coordination \cite{my_ccce, my_rrccce}}

In realistic ATM settings, stakeholders’ cost functions are uncertain and privately observed. 
If coordination is designed using nominal utilities, profitable deviations may arise once realized costs differ from the coordinator’s estimate. 
To address this issue, we extend correlated equilibrium to a probabilistic setting and introduce a chance-constrained correlated equilibrium formulation that explicitly accounts for payoff uncertainty.

\begin{definition}[Chance-constrained correlated equilibrium]
Assume utilities are subject to stochastic perturbations, $u_i = \bar u_i + \eta_i$, where $\eta_i \sim N(0,\sigma_i^2)$. 
A distribution $z \in \Delta(A)$ is a chance-constrained correlated equilibrium with confidence level $\alpha \in [0,1]$ if, for every player $i$ and any actions $a_i,a_i' \in A_i$,
\begin{equation}
\small{\textstyle{
\mathbb{P}\!\left(
\sum_{a_{-i}} z(a_i,a_{-i})
\big( u_i(a_i,a_{-i}) - u_i(a_i',a_{-i}) \big)
\ge 0
\right) \ge \alpha.}}
\end{equation}
\end{definition}

The confidence level $\alpha$ introduces an explicit robustness–efficiency trade-off: larger $\alpha$ reduces deviation risk but restricts the feasible coordination set. 
In collaborative virtual queue coordination experiments with uncertain airline delay costs, the proposed mechanism reduces accumulated delay relative to first-come-first-served baselines while significantly limiting realized deviations.

\subsection{Mechanism design for noncooperative coordination}

While correlated signaling aligns actions through centralized coordination, achieving consensus in a fully decentralized setting remains a fundamental challenge. 
% For example, when multiple feasible equilibria exist, self-interested agents may fail to coordinate on a common outcome without an explicit decentralized mechanism.

\begin{figure}[t!]
    \centering
    \includegraphics[width=0.85\linewidth]{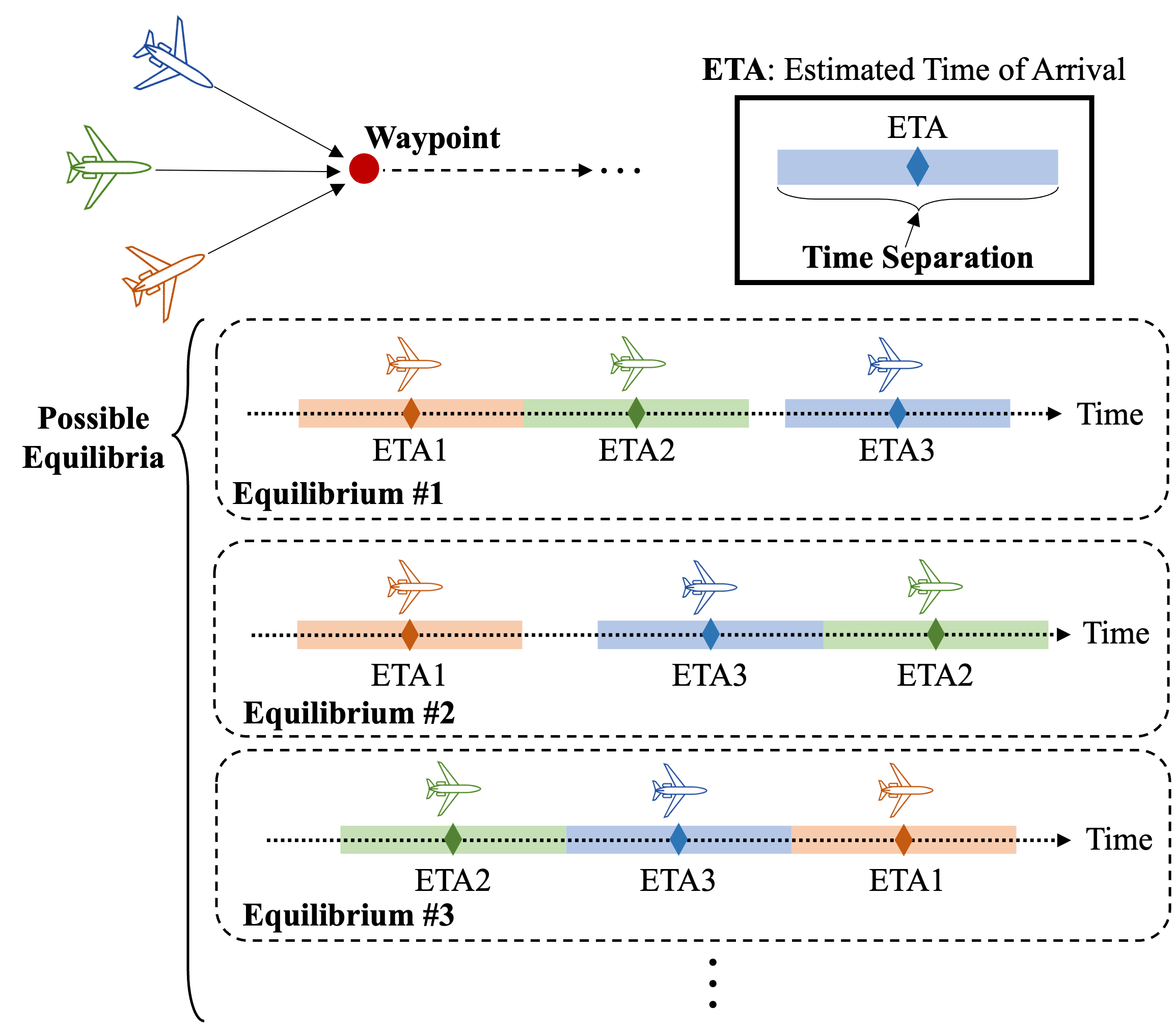}
    \caption{Equilibrium selection in decentralized arrival sequencing. 
    Different arrival orderings satisfy time-separation constraints, creating multiple feasible equilibria with conflicting airline preferences. 
    Trading auction for Consensus (TACo) enables incentive-compatible selection without centralized enforcement.}
    \label{fig:taco}
\end{figure}

\subsubsection{Trading-based equilibrium selection (TACo) \cite{my_taco}}

To achieve decentralized consensus without centralized authority, we developed the \emph{Trading Auction for Consensus} (TACo). 
TACo formulates equilibrium selection as a trading-asset allocation problem. Agents exchange transferable secondary assets (e.g., credits) to reconcile conflicting preferences.

Rather than enforcing a particular outcome, TACo reshapes incentives so that a mutually preferred equilibrium becomes individually rational. 
The mechanism operates through broadcast communication, requires no direct pairwise negotiation, and preserves private valuation information. We establish three properties:
\begin{enumerate}
    \item Finite-time convergence to a consensus equilibrium,
    \item Preservation of private valuation information,
    \item Incentive compatibility under self-interested behavior.
\end{enumerate}

In decentralized arrival sequencing and coordination games, illustrated in \Cref{fig:taco}, TACo selects socially efficient and fair equilibria while ensuring stakeholders' incentive compatibility, compared to randomized equilibrium selection or voting-based schemes.

\subsubsection{Iterative negotiation with oversight \cite{my_iteartive}}

While TACo enables decentralized consensus, safety-critical ATM applications require explicit system-level guarantees. 
We therefore developed an iterative negotiation framework with lightweight regulatory oversight.

Agents iteratively update strategies through local negotiation, while an authority adjusts a taxation parameter $\kappa$ based on reserve shortfalls. 
Rather than enforcing actions, oversight reshapes incentives, steering the system toward feasible equilibria without overriding decentralized decision-making. 
We prove finite-round termination and derive bounds linking intervention level, $\kappa\in[0,1]$, to both convergence speed and efficiency gap, thereby quantifying the autonomy–performance trade-off.

We demonstrate the framework on a decentralized formulation of the \emph{collaborative trajectory options program} (CTOP), where multiple airspace managers independently generate trajectory candidates under shared constraints. 
Experiments show that increasing $\kappa$ strengthens coordination, improves system efficiency and fairness, and approaches centralized performance at the cost of additional negotiation rounds.
This framework formalizes a middle ground between full decentralization and centralized control: agents retain autonomy, while limited intervention ensures system-level guarantees.

\subsection{Noncooperative dynamics and equilibrium steering}

Static equilibrium concepts alone do not guarantee desirable system-level outcomes in decentralized ATM. 
Even when equilibria exist, decentralized best-response updates may converge to inefficient or congested configurations. 
To address this, we explicitly analyze noncooperative dynamics induced by best-response behavior.

\begin{figure}[t!]
    \centering
    \includegraphics[width=0.85\linewidth]{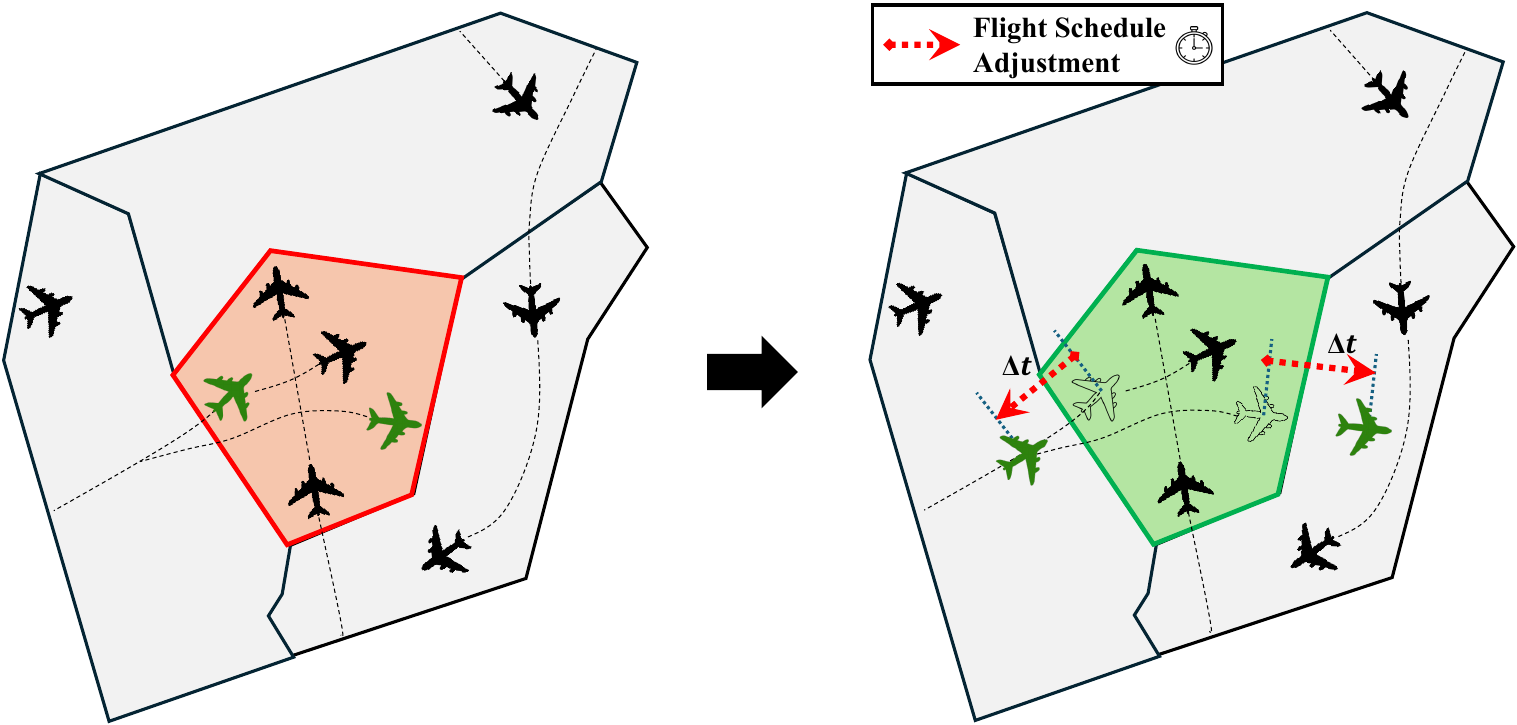}
    \caption{Decentralized sector overload mitigation through schedule adjustment. 
    Left: a sector experiences congestion due to overlapping flight trajectories. 
    Right: decentralized schedule shifts ($\Delta t$) redistribute traffic temporally, alleviating overload without centralized enforcement.}
    \label{fig:brd}
\end{figure}

\subsubsection{Noncooperative dynamics in congestion mitigation \cite{my_brd}}

We study decentralized sector overload mitigation, where multiple congested sectors independently adjust flight departure times to reduce overload. 
Each sector acts as a self-interested agent and updates decisions through best-response dynamics (BRD).

Let $L_i(x)$ denote the overload in sector $i$ under joint schedule profile $x$, defined as the excess demand above sector capacity. 
Each sector minimizes
\begin{equation}
\textstyle{
J_i(x) = L_i(x) + \kappa \sum_{j \neq i} L_j(x),
}
\end{equation}
where $\kappa \in [0,1]$ is a cooperativeness parameter. 
When $\kappa = 0$, sectors minimize only their own overload; when $\kappa = 1$, they fully internalize system-wide congestion.

We prove that the induced game admits potential or ordinal potential structure and that BRD converges to a pure Nash equilibrium in finite time under mild conditions.

Although full cooperation ensures overload elimination, it is not required. 
Monte Carlo simulations using 24-hour European flight data show that even a very small positive $\kappa$—preserving self-prioritization—substantially reduces system overload while maintaining decentralized autonomy. 
Thus, limited cooperation suffices to achieve robust and high-quality decentralized coordination.

\subsubsection{Dynamic incentive design for equilibrium steering}

\begin{figure}[hbt!]
    \centering    \includegraphics[width=0.9\linewidth]{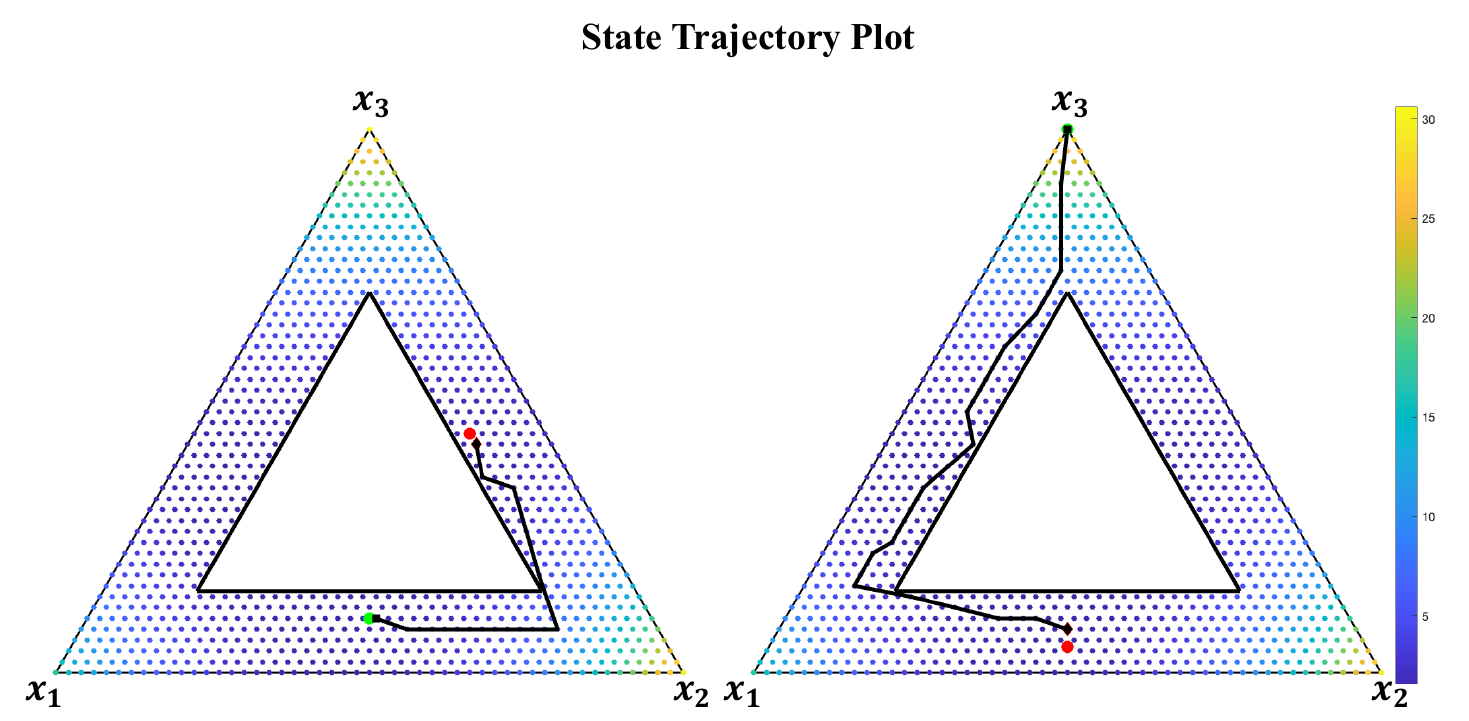}
    \caption{Equilibrium steering in a congestion game. 
    The simplex represents the joint strategy space, and the inner white triangle denotes a non-convex feasibility constraint. 
    From an initial state (green), dynamic incentive signals guide the best-response trajectory toward a desired equilibrium (red) while maintaining feasibility.}
    \label{fig:dynamic}
\end{figure}

When decentralized dynamics converge to inefficient equilibria, structural analysis alone is insufficient. 
We therefore study dynamic incentive design for equilibrium steering.

We model the induced system as
$x_{t+1} = \mathrm{BR}(x_t; u_t)$,
where $u_t$ denotes bounded incentive signals added to agents’ cost functions. 
The objective is to steer the equilibrium trajectory toward a desirable region while maintaining feasibility 
(i.e., $x_t \in \mathcal{X}_{\text{safe}}$, 
$u_t \in \mathcal{U}_{\text{bounded}}$).

This formulation treats incentive design as a predictive control problem over equilibrium dynamics. 
Key questions include equilibrium reachability under bounded intervention, minimal incentive magnitude required to shift equilibrium basins, and safety preservation during transient evolution.

In a congestion game motivated by AAM traffic allocation, illustrated in \Cref{fig:dynamic}, we demonstrate through a toy example that optimal incentive sequences can steer best-response trajectories toward target equilibria while respecting nonlinear feasibility constraints. 
% These preliminary results suggest that dynamic, safety-aware incentive planning can expand the set of achievable equilibria beyond static mechanism design.

\section{Conclusion}

We establish a unified view of decentralized air traffic management as a noncooperative system whose equilibria can be engineered, selected, and steered through incentive design.
Overall, the proposed dissertation demonstrates that effective coordination in decentralized air traffic systems does not require centralized enforcement. 
Instead, by aligning incentives, shaping equilibrium structure, and leveraging calibrated intervention, system-level performance can emerge from individually rational decision-making.

\bibliographystyle{IEEEtran} 
\bibliography{reference}

@article{atm_problem_1,
  title={Evaluation of tactical conflict resolution algorithms for enroute airspace},
  author={Paielli, Russell A},
  journal={J. Aircr.},
  volume={48},
  number={1},
  pages={324--330},
  year={2011}
}

@article{atm_problem_2_aam,
  title={Small aircraft transportation system, higher volume operations concept and research summary},
  author={Baxley, Brian T and Williams, Daniel and Consiglio, Maria and Adams, Cathy and Abbott, Terrence},
  journal={J. Aircr.},
  volume={45},
  number={6},
  pages={1825--1834},
  year={2008}
}

@inproceedings{atm_central,
  title={A safety-driven approach to exploring and comparing air traffic management concepts for enabling urban air mobility},
  author={Poh, Justin and Leveson, Nancy G and Neogi, Natasha A},
  booktitle={Int. Conf. Res. Air Transp. (ICRAT)},
  year={2024}
}

@article{atm_central_2,
  title={Decentralized control synthesis for air traffic management in urban air mobility},
  author={Bharadwaj, Suda and Carr, Steven and Neogi, Natasha and Topcu, Ufuk},
  journal={IEEE Trans. Control Netw. Syst.},
  volume={8},
  number={2},
  pages={598--608},
  year={2021},
  publisher={IEEE}
}

@inproceedings{cent_resilience_prob,
  title={Urban Air Mobility: A Review of Recent Advances in Communication, Management, and Sustainability},
  author={He, Zhitong and Wang, Zijing and Li, Lingxi},
  booktitle={Int. Conf. Cyber-Phys. Soc. Intell. (CPSI)},
  pages={1--6},
  year={2025},
  organization={IEEE}
}

@article{cent_resilience_prob_2,
  title={A data-driven approach to resilience in air traffic management: case study Barcelona area control centre},
  author={Mirkovic, Bojana and Timotic Petkovic, Doroteja and Netjasov, Fedja and Crnogorac, Dusan and Verdonk Gallego, Christian Eduardo and Xia, Chen and Malakis, Stathis},
  journal={Cogn. Technol. Work},
  volume={26},
  number={3},
  pages={457--485},
  year={2024},
  publisher={Springer}
}

@inproceedings{decent_intro,
  title={Evolutionary concepts for decentralized air traffic flow management},
  author={Adams, Milton and Kolitz, Stephan and Adani, Amedeo},
  booktitle={Guid. Navig. Control Conf.},
  pages={3857},
  year={1997}
}

@inproceedings{decent_intro_2,
  title={Distributed air-ground traffic management for en route flight operations},
  author={Green, Steven and Bilimoria, Karl and Ballin, Mark},
  booktitle={AIAA Guid. Navig. Control Conf. Exhib.},
  pages={4064},
  year={2001}
}

@inproceedings{prev_decent,
  title={NASA Langley and NLR research of distributed air/ground traffic management},
  author={Ballin, Mark and Hoekstra, Jacco and Wing, David and Lohr, Gary},
  booktitle={AIAA Aircr. Technol. Integr. Oper. (ATIO) Tech. Forum},
  pages={5826},
  year={2002}
}

@inproceedings{prev_decent_block,
  title={Decentralizing air traffic flow management with blockchain-based reinforcement learning},
  author={Duong, Ta and Todi, Ketan Kumar and Chaudhary, Umang and Truong, Hong-Linh},
  booktitle={IEEE Int. Conf. Ind. Inf. (INDIN)},
  volume={1},
  pages={1795--1800},
  year={2019},
  organization={IEEE}
}

@article{prev_decent_nash,
  title={Decentralized free-flow traffic management based on nash equilibrium},
  author={Im, Jaehan and Ahn, Jaemyung},
  journal={J. Aerosp. Inf. Syst.},
  volume={20},
  number={4},
  pages={195--203},
  year={2023},
  publisher={American Institute of Aeronautics and Astronautics}
}

@inproceedings{prev_decent_utm,
  title={Fairness in decentralized strategic deconfliction in UTM},
  author={Evans, Antony D and Egorov, Maxim and Munn, Steven},
  booktitle={AIAA Scitech Forum},
  pages={2203},
  year={2020}
}

@inproceedings{prev_decent_utm_2,
  title={Unmanned aircraft system traffic management (UTM) concept of operations},
  author={Kopardekar, Parimal and Rios, Joseph and Prevot, Thomas and Johnson, Marcus and Jung, Jaewoo and Robinson, John E},
  booktitle={AIAA Aviat. Forum Expo.},
  number={ARC-E-DAA-TN32838},
  year={2016}
}

@inproceedings{prev_decent_utm_hier,
  title={A safety-driven approach to exploring and comparing air traffic management concepts for enabling urban air mobility},
  author={Poh, Justin and Leveson, Nancy G and Neogi, Natasha A},
  booktitle={Int. Conf. Res. Air Transp. (ICRAT)},
  year={2024}
}

@inproceedings{tomlin,
  title={Noncooperative conflict resolution [air traffic management]},
  author={Tomlin, Claire and Pappas, George J and Sastry, Shankar},
  booktitle={Proc. IEEE Conf. Decis. Control},
  volume={2},
  pages={1816--1821},
  year={1997},
  organization={IEEE}
}

@article{noncoop_required,
  title={HF002: Applied Game Theory to Enhance Air Traffic Control Training},
  author={Barker, Kash and Lakshmivarahan, S and Ghorbani-Renani, Nafiseh and Rangrazjeddi, Alireza and Gonz{\'a}lez, Andr{\'e}s D and Wood, Robert and Demagalski, Jason}
}

@article{my_iteartive,
  title={Iterative Negotiation and Oversight: A Case Study in Decentralized Air Traffic Management},
  author={Im, Jaehan and Clarke, John-Paul and Topcu, Ufuk and Fridovich-Keil, David},
  journal={arXiv preprint arXiv:2511.17625},
  year={2025}
}

@article{my_rrce,
  title={Coordination in noncooperative multiplayer matrix games via reduced rank correlated equilibria},
  author={Im, Jaehan and Yu, Yue and Fridovich-Keil, David and Topcu, Ufuk},
  journal={IEEE Control Syst. Lett.},
  volume={8},
  pages={1637--1642},
  year={2024},
  publisher={IEEE}
}

@article{my_taco,
  title={Noncooperative Consensus via a Trading-based Auction},
  author={Im, Jaehan and Fotiadis, Filippos and Delahaye, Daniel and Topcu, Ufuk and Fridovich-Keil, David},
  journal={arXiv preprint arXiv:2502.03616},
  year={2025}
}

@article{my_brd,
  title={Game-theoretic Decentralized Coordination for Airspace Sector Overload Mitigation},
  author={Im, Jaehan and Delahaye, Daniel and Fridovich-Keil, David and Topcu, Ufuk},
  journal={arXiv preprint arXiv:2511.13770},
  year={2025}
}

@article{brugnara2023market,
  title={A market mechanism for multiple air traffic resources},
  author={Brugnara, Irene and Castelli, Lorenzo and Pesenti, Raffaele},
  journal={Transportation Research Part E: Logistics and Transportation Review},
  volume={178},
  pages={103255},
  year={2023},
  publisher={Elsevier}
}

@article{my_ccce,
  title={Chance-Constrained Correlated Equilibria for Robust Noncooperative Coordination},
  author={Im, Jaehan and Topcu, Ufuk and Fridovich-Keil, David},
  journal={arXiv preprint arXiv:2603.14141},
  year={2026}
}

@article{my_rrccce,
  title={Scalable Coordination with Chance-Constrained Correlated Equilibria via Reduced-Rank Structure},
  author={Im, Jaehan and Fridovich-Keil, David and Topcu, Ufuk},
  journal={arXiv preprint arXiv:2604.00456},
  year={2026}
}

@article{my_security,
  title={Secure Coordination for Vertiport Sequencing in Advanced Air Mobility},
  author={Im, Jaehan and Fotiadis, Filippos and Topcu, Ufuk and Fridovich-Keil, David},
  journal={arXiv preprint arXiv:2605.21771},
  year={2026}
}

\end{document}